\documentstyle[12pt,epsf]{ioplppt} 

\begin{document} 
\jl{3}
\comment{
Comment on ``Boson-fermion model beyond the mean-field approximation''
}[Comment on ``Boson-fermion model...'']
\author{R Friedberg\dag, H C Ren\ddag, 
and O Tchernyshyov\dag\ftnote{3}{E-mail: olegt@cuphyb.phys.columbia.edu}}
\address{
\dag\ Department of Physics, Columbia University, 
New York, New York 10027}
\address{\ddag\ Department of Physics, The Rockefeller University, 
New York, New York 10021}

\begin{abstract}
In a recent paper [Alexandrov A S 1996 \JPCM {\bf 8} 6923--32; 
cond-mat/9603111], it has been suggested that 
there is no Cooper pairing in boson-fermion models of 
superconductivity.  We show that this conjecture is based on an 
inconsistent approximation that violates an exact identity.  
Quite generally, the divergence of the fermion t-matrix (the Thouless 
criterion) is accompanied by the condensation of a boson mode.  
\end{abstract}

\pacs{74.20, 11.25.Db}

\maketitle

\nosections

In a recent publication \cite{Alexandrov}, Alexandrov has argued that
there is no Cooper pairing in boson-fermion models \cite{FL,Ranninger}.
It was suggested that the Thouless criterion for the onset
of long-range order is inconsistent with the requirement that the
physical energy of a boson be non-negative.  

We would like to point out in this respect that the proof 
is based on two approximations and therefore is not exact.  
(a) The fermion t-matrix is approximated by a ladder series.  
(b) The boson self-energy is approximated by a single two-fermion bubble.  
Below we will show that there exists an exact identity relating these two 
quantities.  Therefore, choosing an approximation for the t-matrix 
fixes the form of the approximate boson propagator.  Alexandrov's
choice (b) violates that identity and makes his approximation 
inconsistent (in the sense of Baym and Kadanoff \cite{Baym}).  

\begin{figure}[thb]
\begin{center}
\leavevmode
\epsfxsize 5.5in
\epsffile{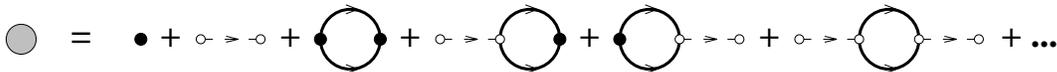}
\end{center}
\caption{The series for the fermion t-matrix (shaded circle) 
in Alexandrov's approximation,
through second order in $V\equiv V_c+v^2D_0({\bf q},\Omega_n)$. 
Open dot: boson-fermion coupling $v$.  
Filled dot: on-site repulsion $V_c$.
Thick solid line: dressed fermion propagator $G({\bf k},\omega_n)$.
Thin broken line: bare boson propagator $D_0({\bf q},\Omega_n)$.}
\label{t-matrix}
\end{figure}

Although Equation 7 in \cite{Alexandrov}, which determines $T_c$, 
is derived from the linearized gap equation, an alternative
(but fully equivalent) derivation uses the divergence of the
fermion t-matrix at zero energy and momentum as a criterion
for the transition \cite{Thouless}.  Alexandrov adopts an effective 
short-range fermion interaction in the form $V = v^2 D_0(0,0) + V_c$
(the first two diagrams in Figure \ref{t-matrix}),
where $D_0({\bf q},\Omega_n)$ is the free boson propagator, 
$v$ is the boson-fermion coupling, and $V_c$ is a short-range
fermion repulsion. 
In the ladder approximation, one obtains the 
usual result for the t-matrix:
\begin{equation}
T(0,0) = \frac{V}{1-V{\cal B}},
\end{equation}
where, in Alexandrov's notation, ${\cal B} 
= - N^{-1}\int {\rm d} p\ G(p)G(-p)$ 
is a short-hand for a bubble with two fermion lines.
The divergence of the t-matrix then requires that
\begin{equation}
\label{divergence}
1-[v^2 D_0(0,0) + V_c]{\cal B} = 0,
\end{equation}
which is precisely Equation 7 of Alexandrov.

\begin{figure}[h!]
\begin{center}
\leavevmode
\epsfxsize 3.0in
\epsffile{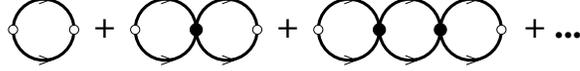}
\end{center}
\caption{Boson self-energy in the ladder approximation.  Only the first graph
is included by Alexandrov.}
\label{ladder}
\end{figure}

Turning to the boson self-energy, we supplement Alexandrov's single diagram 
(a bubble with two fermion lines) by a ladder series to account for 
Coulomb repulsion $V_c$ between the intermediate fermions (Figure 
\ref{ladder}). This choice will be justified below.   
Upon including these diagrams, the boson propagator at ${\bf q}=0$ 
and $\Omega_n=0$ is given by the equation
\begin{equation}
D^{-1}(0,0) = D_0^{-1}(0,0) - \frac{v^2{\cal B}}{1-V_c{\cal B}},
\end{equation}
At the condensation temperature, bosons with zero momentum have zero energy, 
i.e., $D(0,0)=\infty$, and we recover Equation \ref{divergence}.  

\begin{figure}[t]
\begin{center}
\leavevmode
\epsfxsize 4in
\epsffile{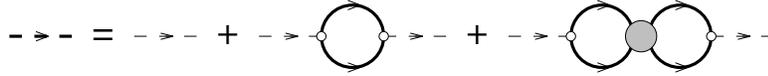}
\end{center}
\caption{An identity for the exact boson propagator (thick dashed line).
Thin dashed line: free boson propagator.}
\label{identity}
\end{figure}

Evidently, when the boson self-energy is properly modified, 
the Thouless criterion no longer contradicts the 
sum rule (6) in \cite{Alexandrov}.  Instead, we find that {\em the t-matrix
diverges exactly at the condensation temperature of the
bosons.}  This remarkable conclusion is substantiated by
an {\em exact} identity presented in Figure \ref{identity}.  
It indicates that the divergence of $T(0,0)$ {\em must} be accompanied by 
that of $D(0,0)$.

The proof of the identity uses the equations of motion for 
a non-self-interacting boson field $\phi(x,t)$ with a Hamiltonian 
\begin{equation}
H_B = \int {\rm d}x\ \phi^\dagger(x,t){\cal H}_B(x)\phi(x,t),
\end{equation}
where ${\cal H}_B(x)$ is a linear operator acting on $x$.  The 
boson field is coupled to a fermion field $\psi_\sigma(x,t)$ by
the interaction Hamiltonian
\begin{equation}
H_{BF} = v\int {\rm d}x\ 
[\phi^\dagger(x,t)\psi_\uparrow(x,t)\psi_\downarrow(x,t) + \mbox{ H.c.}].
\end{equation}
In what follows, $x$ denotes spatial coordinates and the (complex) time 
$t$ varies along a straight line between 0 and $\tau=-{\rm i}\hbar/k_{\rm B}T$.

Introduce time-ordered boson ($D$), two-fermion ($G_2$) and mixed ($M$) 
propagators 
\begin{eqnarray}
D(x,t;x',t') 
= -{\rm i}\langle T[\phi(x,t)\phi^\dagger(x',t')]\rangle,\\
M(x,y,t;x',t') 
= -{\rm i}\langle T[\psi_\uparrow(x,t)\psi_\downarrow(y,t)
\phi^\dagger(x',t')]\rangle,\\
G_2(x,y,t;x',y',t') 
= -{\rm i}\langle T[\psi_\uparrow(x,t)\psi_\downarrow(y,t)
\psi_\downarrow^\dagger(y',t')\psi_\uparrow^\dagger(x',t')]\rangle.
\end{eqnarray}
By using equations of motions for $\phi$ and $\phi^\dagger$, we obtain
\begin{eqnarray}
[{\rm i}\partial/\partial t - {\cal H}_B(x)] D(x,t;x',t') 
= \delta(x-x')\delta(t-t') + v M(x,x,t;x',t'),\\ 
{}[-{\rm i}\partial/\partial t' - {\cal H}_B(x')] M(x,x,t;x',t') 
= v G_2(x,x,t;x',x',t').  
\end{eqnarray}
These differential relations can be integrated with the aid of 
the free ($v=0$) boson propagator $D_0(x,t;x',t')$.  The following 
identity is then obtained for the corresponding Fourier coefficients:
\begin{eqnarray}
\fl D(x;x'|n) = D_0(x;x'|n) \nonumber \\ 
+ v^2\int {\rm d}x_1 \int {\rm d}x_2\  
D_0(x;x_1|n) G_2(x_1,x_1;x_2,x_2|n) D_0(x_2;x'|n).
\label{theidentity}
\end{eqnarray}
$D(x;x'|n)$ is defined in a standard way,
\begin{equation}
D(x,t;x',t') = \frac{1}{\tau}\sum_{n=-\infty}^{\infty}D(x;x'|n)
\exp{[-{\rm i}\Omega_n(t-t')]},
\end{equation}
where $\Omega_n=2\pi n/\tau$.
Finally, we isolate the disconnected part of $G_2$ ($GG$) and 
express the connected part in terms of the proper vertex ($GGTGG$), as 
shown in Figure \ref{identity}.  

\begin{figure}[bt]
\begin{center}
\leavevmode
\epsfxsize 5in
\epsffile{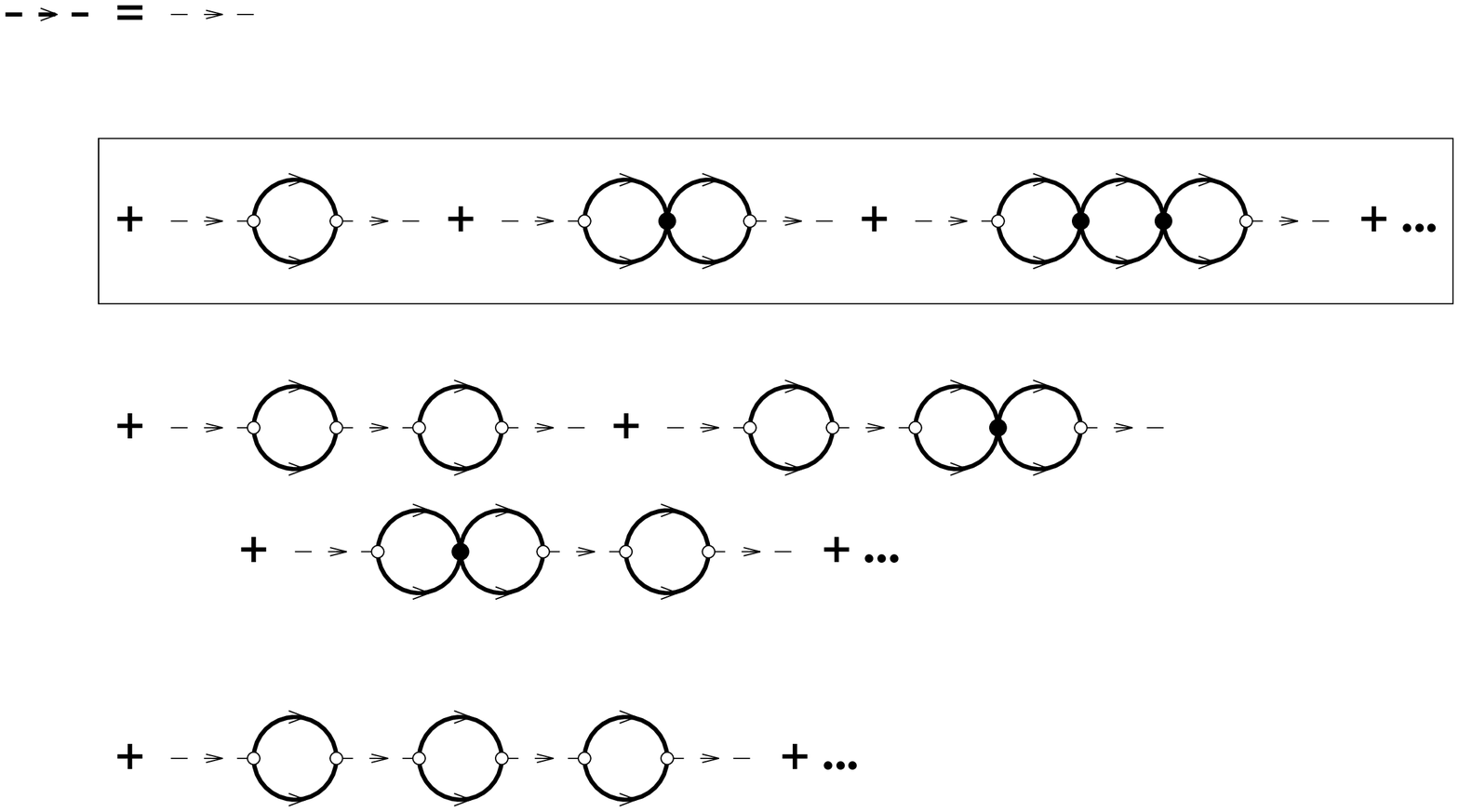}
\end{center}
\caption{The boson propagator obtained in a self-consistent approximation
(by combining Figures 1 and 3).  The diagrams are grouped
according to the number of free boson lines.  The boxed set of graphs
(with the external boson lines amputated) gives the boson self-energy.}
\label{boson}
\end{figure}

We are now in position to justify our choice of Feynman diagrams for the
approximate boson self-energy (Figure \ref{ladder}).  To this end, we can
determine the dressed boson propagator directly from the identity 
of Figure \ref{identity}, by using the approximate expression for the 
t-matrix (Figure \ref{t-matrix}) as the input.  In this way, 
the exact result (\ref{theidentity}) is built into the approximate theory 
from the start.  The resulting boson propagator is shown in Figure 
\ref{boson}.  Higher-order graphs replaced with the dots are generated 
by similarly omitted graphs of Figure \ref{t-matrix}.  The diagrams with
only two free boson lines determine the boson self-energy, which by 
inspection coincides with our choice (Figure \ref{ladder}).

In conclusion, we have shown that Alexandrov's conjecture
(no Cooper pairing in boson-fermion models) is incorrect.  A 
careful revision of his approach reveals an exact identity 
that expresses equivalence of the Thouless criterion to
Bose condensation in a wide class of boson-fermion models.
Being non-self-consistent, the approximation made by Alexandrov 
violates this identity and therefore leads to a contradiction.

\end{document}